\documentclass[a4paper]{jpconf}
\usepackage{graphicx}
\usepackage{bm}
\usepackage{epsfig}
\usepackage{graphics}
\usepackage{amsmath}
\usepackage{amssymb}
\usepackage{latexsym}

\def\beq{\begin{equation}}
\def\eeq{\end{equation}}
\def\bea{\begin{eqnarray}}
\def\eea{\end{eqnarray}}

\newcommand{\Eref}[1]{Eq.~(\ref{#1})}
\renewcommand{\Sref}[1]{Sec.~\ref{#1}}
\renewcommand{\Fref}[1]{Fig.~\ref{#1}}
\renewcommand{\Tref}[1]{Table~\ref{#1}}
\newcommand{\cref}[1]{Ref.~\cite{#1}}
\newcommand{\Erefs}[2]{Eqs.~(\ref{#1}) and (\ref{#2})}
\newcommand{\beqs}{\begin{subequations}}
\newcommand{\eeqs}{\end{subequations}}

\def\lf{\left(}
\def\rg{\right)}
\renewcommand{\etal}{{\it et al.}}
\newcommand{\GeV}{{\mbox{\rm GeV}}}

\def\mcr{{\tt micrOMEGAs}}

\newcommand{\bmm}{{\ensuremath{{\rm BR}\lf B_s\to \mu^+\mu^-\rg}}}
\newcommand{\bsg}{{\ensuremath{{\rm BR}\lf b\to s\gamma\rg}}}
\newcommand{\btn}{{\ensuremath{{\rm R}\lf B_u\to \tau\nu\rg}}}
\newcommand{\Dam}{{\ensuremath{\delta a_{\mu}}}}
\newcommand{\Omx}{{\ensuremath{\Omega_{\tilde\chi} h^2}}}

\newcommand{\mx}{{\ensuremath{m_{\tilde\chi}}}}
\newcommand{\Dst}{{\ensuremath{\Delta_{\tilde\tau_2}}}}

\newcommand{\Mg}{{\ensuremath{M_{1/2}}}}
\newcommand{\AMg}{{\ensuremath{A_0/M_{1/2}}}}
\newcommand{\xx}{{\ensuremath{\tilde\chi}}}

\begin{document}

\title{Dark Matter and Higgs Mass in the CMSSM with Yukawa 
Quasi-Unification}


\author{N Karagiannakis$^1$, G Lazarides$^1$,
C Pallis$^{2}$}

\address{$^1$ Physics Division, School of
Technology, Aristotle University of Thessaloniki, Thessaloniki
54124, Greece}
\address{$^2$ Department of Physics, University
of Cyprus, P.O. Box 20537, CY-1678 Nicosia, CYPRUS}

\ead{nikar@auth.gr, lazaride@eng.auth.gr, kpallis@auth.gr}

\date{\today}

\begin{abstract}
We present an updated analysis of the constrained minimal supersymmetric 
standard model with $\mu>0$ supplemented by an `asymptotic' Yukawa coupling 
quasi-unification condition, which allows an acceptable $b$-quark mass. 
Imposing constraints from the cold dark matter abundance in the universe, 
$B$ physics, the muon anomalous magnetic moment, and the mass $m_h$ of the 
lightest neutral CP-even Higgs boson, we find that the lightest neutralino 
cannot act as a cold dark matter candidate. This is mainly because the 
upper bound on the lightest neutralino relic abundance from cold dark 
matter considerations, despite the fact that this abundance is drastically 
reduced by neutralino-stau coannihilations, is incompatible with the recent 
data on the branching ratio of $B_s\to\mu^+\mu^-$. Allowing for a different 
particle, such as the axino or the gravitino, to be the lightest 
supersymmetric particle and, thus, constitute the cold dark matter in the 
universe, we find that the predicted $m_h$'s in our model favor the range 
$(119-126)~\GeV$.

\end{abstract}

\section{Introduction}\label{sec:intro}

The well-known \emph{constrained minimal supersymmetric standard
model} (CMSSM) \cite{Cmssm0,Cmssm,cmssm1,cmssm2}, which is a highly
predictive version of the \emph{minimal supersymmetric standard
model} (MSSM) based on universal boundary conditions for the soft
\emph{supersymmetry} (SUSY) breaking parameters, can be further
restricted by being embedded in a SUSY \emph{grand unified theory}
(GUT) with a gauge group containing $SU(4)_c$ and $SU(2)_R$. This
can lead \cite{pana} to `asymptotic' \emph{Yukawa unification}
(YU)~\cite{als}, i.e. the exact unification of the third
generation Yukawa coupling constants $h_t$, $h_b$, and $h_\tau$ of
the top quark, the bottom quark, and the tau lepton, respectively,
at the SUSY GUT scale $M_{\rm GUT}$. The simplest GUT gauge group
which contains both $SU(4)_c$ and $SU(2)_R$ is the
\emph{Pati-Salam} (PS) group $G_{\rm PS}=SU(4)_c\times
SU(2)_L\times SU(2)_R$ \cite{leontaris,jean} -- for YU within
$SO(10)$, see Refs.~\cite{baery,raby}.

However, given the experimental values of the top-quark and
tau-lepton masses (which, combined with YU, naturally restrict
$\tan\beta\sim50$), the CMSSM supplemented by the assumption of YU
yields unacceptable values of the $b$-quark mass $m_b$ for both
signs of the parameter $\mu$. This is due to the presence of
sizable SUSY corrections \cite{copw} to $m_b$ (about 20$\%$),
which arise \cite{copw,pierce} from sbottom-gluino (mainly) and
stop-chargino loops and have the sign of $\mu$ -- with the
standard sign convention of Ref.~\cite{sugra}. The predicted
tree-level $m_b(M_Z)$, which turns out to be close to the upper
edge of its $95\%$ \emph{confidence level} (c.l.) experimental
range, receives, for $\mu>0$ [$\mu<0$], large positive [negative]
corrections which drive it well above [a little below] the allowed
range. Consequently, for both signs of $\mu$, YU leads to an
unacceptable $m_b(M_Z)$ with the $\mu<0$ case being much less
disfavored.

The usual strategy to resolve this discrepancy is the introduction
of several kinds of nonuniversalities in the scalar \cite{baery,
raby} and/or gaugino \cite{shafi,nath} sector of MSSM with an
approximate preservation of YU. On the contrary, in
Ref.~\cite{qcdm}, concrete SUSY GUT models based on the PS gauge
group are constructed which naturally yield a moderate deviation
from exact YU and, thus, can allow acceptable values of the
$b$-quark mass for both signs of $\mu$ within the CMSSM. In
particular, the Higgs sector of the simplest PS model
\cite{leontaris, jean} is extended so that the electroweak Higgs
fields are not exclusively contained in a $SU(2)_L\times SU(2)_R$
bidoublet superfield, but receive subdominant contributions from
other representations too. As a consequence, a moderate violation
of YU is naturally obtained, which can allow an acceptable
$b$-quark mass even with universal boundary conditions. It is also
remarkable that the resulting extended SUSY PS models support new
successful versions \cite{axilleas} of hybrid inflation based
solely on renormalizable superpotential terms.

These models provide us with a set of `asymptotic' Yukawa
quasi-unification conditions, which replace exact YU and can be
applicable \cite{qcdm,muneg,nova} for both signs of the MSSM parameter
$\mu$. We focus here on the $\mu>0$ case since $\mu<0$ is strongly
disfavored by the constraint arising from the deviation $\delta
a_\mu$ of the measured value of the muon anomalous magnetic moment
$a_\mu$ from its predicted value $a^{\rm SM}_\mu$ in the
\emph{standard model} (SM). Indeed, $\mu<0$ is defended only at 
3$\sigma$ by the calculation of $a^{\rm SM}_\mu$ based on the 
$\tau$-decay data which is presented in Ref.~\cite{g2davier}, 
whereas there is a stronger and stronger tendency at present to 
prefer the $e^+e^-$-annihilation data for the calculation of 
$a^{\rm SM}_\mu$, which favor the $\mu>0$ regime. Moreover, 
in \cref{Jen}, it was claimed that, after some improvements, the 
$\tau$-based result shifts considerably towards the $e^+e^-$-based
one.

Let us recall that, in this case, the suitable `asymptotic' Yukawa
quasi-unification condition applied \cite{qcdm,muneg,nova} is
\begin{equation}
h_t:h_b:h_\tau=|1+c|:|1-c|:|1+3c|. \label{minimal}
\end{equation}
This relation depends on a single parameter $c$, which is taken,
for simplicity, to be real and lying in the range $0<c<1$. With
fixed masses for the fermions of the third generation, we can
determine the parameters $c$ and $\tan\beta$ so that
Eq.~(\ref{minimal}) is satisfied. In contrast to the original
version of the CMSSM \cite{Cmssm, cmssm1, cmssm2}, therefore,
$\tan\beta$ is not a free parameter, but can be restricted,
within our set-up, via \Eref{minimal} to relatively large values.
The remaining free parameters of our model are the universal soft
SUSY breaking parameters defined at $M_{\rm GUT}$, i.e.
\begin{equation}
M_{1/2},~~m_0,~~\mbox{and}~~A_0, \label{param}
\end{equation}
where the symbols above denote the common gaugino mass, scalar
mass, and trilinear scalar coupling constant, respectively. These
parameters can be restricted by employing a number of experimental
and cosmological requirements as in Refs.~\cite{qcdm,muneg,nova} and
most recently in Refs.~\cite{yqu,shafi11}. In this talk, we review 
the results of \cref{yqu} implementing the following improvements:

\begin{itemize}

\item We do not take into account the upper bound on $\mx$ 
implied by the lower bound on $\delta a_\mu$ from the $\tau$-based 
calculation of Ref.~\cite{g2davier} raising, thereby, the upper 
bound on $\mx$ from the muon anomalous magnetic moment -- see 
\Sref{sec:pheno}. 

\item We employ the recently released data on the branching ratio
of $B_s\to\mu^+\mu^-$ \cite{bmmexp1} and the mass $m_h$ of the 
lightest CP-even Higgs boson \cite{Hlhc}. As a consequence, our
predictions in \cref{yqu} for $m_h$ and the role of the lightest
neutralino as {\it cold dark matter} (CDM) particle have been 
significantly altered.

\end{itemize}

All the cosmological and phenomenological requirements which we
considered in our investigation are exhibited in detail in
Sec.~\ref{sec:pheno}. Restrictions on the parameter space of our
model are derived in Sec.~\ref{results} and our conclusions are
summarized in \Sref{con}.

\section{Cosmological and Phenomenological Constraints}
\label{sec:pheno}

In our investigation, we integrate the two-loop renormalization
group equations for the gauge and Yukawa coupling constants and
the one-loop ones for the soft SUSY breaking parameters between
$M_{\rm GUT}$ and a common SUSY threshold $M_{\rm SUSY}
\simeq(m_{\tilde t_1}m_{\tilde t_2})^{1/2}$ ($\tilde t_{1,2}$ are
the stop mass eigenstates) determined in consistency with the SUSY
spectrum. At $M_{\rm SUSY}$, we impose radiative electroweak
symmetry breaking, evaluate the SUSY spectrum employing the
publicly available calculator {\tt SOFTSUSY} \cite{Softsusy}, and
incorporate the SUSY corrections to the $b$ and $\tau$ mass
\cite{pierce}. The corrections to the $\tau$-lepton mass $m_\tau$
(almost 4$\%$) lead \cite{qcdm,muneg} to a small decrease of
$\tan\beta$. From $M_{\rm SUSY}$ to $M_Z$, the running of gauge
and Yukawa coupling constants is continued using the SM
renormalization group equations.

The parameter space of our model can be restricted by using a
number of phenomenological and cosmological constraints.  We
calculate them using the latest version of the publicly available
code {\tt micrOMEGAs} \cite{micro}. We now briefly discuss these
requirements -- for similar recent analyses, see Ref.~\cite{lhc}
for the CMSSM or Refs.~\cite{shafi,baerlhc} for the MSSM with YU.

\paragraph{SM Fermion Masses.} The masses of the
fermions of the third generation play a crucial role in the
determination of the evolution of the Yukawa coupling constants.
For the $b$-quark mass, we adopt as an input parameter in our
analysis the $\overline{\rm MS}$ $b$-quark mass, which at
1$\sigma$ is \cite{pdata}
\beq m_b \lf m_b\rg^{\overline{\rm MS}}=4.19^{+0.18}_{-0.06}~\GeV.
\eeq
This range is evolved up to $M_Z$ using the central value
$\alpha_s(M_Z)=0.1184$ \cite{pdata} of the strong fine structure
constant at $M_Z$ and then converted to the ${\rm \overline{DR}}$
scheme in accordance with the analysis of Ref.~\cite{baermb}. We
obtain, at $95\%$ c.l.,
\beq 2.745\lesssim  m_b(M_Z)/{\rm GeV}\lesssim 3.13
\label{mbrg}\eeq
with the central value being $m_b(M_Z)=2.84~\GeV$. For the
top-quark mass, we use the central pole mass ($M_t$) as an input
parameter \cite{mtmt}:
\beq M_t=173~\GeV~~\Rightarrow~~m_t(m_t)=164.6~\GeV\eeq
with $m_t(m_t)$ being the running mass of the $t$ quark. We also
take the central value $m_{\tau}(M_Z) = 1.748~\GeV$ \cite{baermb}
of the ${\overline{\rm DR}}$ tau-lepton mass at $M_Z$.

\paragraph{Cold Dark Matter Considerations.}
\label{phenoa} According to the WMAP results \cite{wmap}, the
$95\%$ c.l. range for the CDM abundance is
\beq \Omega_{\rm CDM}h^2=0.1126\pm0.0072. \label{cdmba}\eeq
In the context of the CMSSM, the lightest neutralino 
$\tilde\chi$ can be the {\it lightest supersymmetric 
particle} (LSP) and, thus, naturally arises as a 
CDM candidate. In this case, the requirement that the 
$\tilde\chi$ relic abundance $\Omx$ does not exceed the 
$95\%$ c.l. upper bound from Eq.~(\ref{cdmba}), 
i.e.
\beq \Omx\lesssim0.12,\label{cdmb}\eeq
strongly restricts the parameter space of the CMSSM. This is 
because
$\Omx$  increases, in general, with $\mx$ and so an upper bound
on $m_{\tilde\chi}$ can be derived from Eq.~(\ref{cdmb}). The
calculation of $\Omx$ in \mcr\ includes accurately thermally
averaged exact tree-level cross sections of all the possible
(co)annihilation processes \cite{cmssm1, cdm}, treats poles
\cite{cmssm2, qcdm, nra} properly, and uses one-loop QCD and 
SUSY QCD corrections \cite{copw, qcdm, microbsg} to the Higgs 
decay widths and couplings to fermions. It should, though, be 
noted that the restrictions induced by \Eref{cdmb} can be 
evaded if we adopt one (or a combination) of the following 
scenarios:

\begin{itemize}

\item The cosmological evolution before \emph{Big Bang
nucleosynthesis} (BBN) deviates from the standard one 
\cite{scn0, scn}. Since $\xx$ within the CMSSM is 
essentially a pure bino, the scenario which fits better 
this case is the low reheat temperature scenario
with the decoupling of $\xx$ occurring before reheating. 
This scenario, however, is disfavored since it requires 
a very low reheat temperature $\sim (1-5)~\GeV$. We 
will, thus, assume that the decoupling of the neutralino 
from the cosmic fluid occurs during the conventional 
radiation dominated era.

\item The lightest neutralino is not the LSP and, 
thus, the relic density of another SUSY particle 
\cite{candidates}, which is the LSP, is to account 
for $\Omega_{\rm CDM}h^2$. This particle could be 
the gravitino \cite{gravitino} or the axino 
\cite{axino, Baerax}. The case of gravitino is tightly 
restricted in the CMSSM due to the BBN constraints 
imposed during the decay of the \emph{lightest ordinary 
supersymmetric particle} (LOSP) to it. On the other 
hand, axino CDM \cite{Baerax} is, in general, possible 
once its mass and the reheat temperature 
are chosen appropriately. In such a case, $\xx$ may play 
the role of the LOSP and can contribute to the non-thermal 
production of the LSP. In particular, its contribution to 
the relic density of the LSP is equal to $\Omx$ times the 
ratio of the LSP mass to $\mx$ and, thus, $\Omx$'s 
exceeding the bound in Eq.~(\ref{cdmb}) can be perfectly 
acceptable.

\end{itemize}

\paragraph{The Branching Ratio $\bsg$ of $b\to s\gamma$.} The most 
recent
experimental world average for ${\rm BR}(b\rightarrow s\gamma)$ is
known \cite{bsgexp} to be $\lf3.52\pm0.23\pm0.09\rg\times10^{-4}$
and its updated SM prediction is $\lf3.15\pm0.23\rg\times10^{-4}$
\cite{bsgSM}. Combining in quadrature the experimental and
theoretical errors involved, we obtain the following constraints
on this branching ratio at $95\%$ c.l.:
\beq 2.84\times 10^{-4}\lesssim \bsg \lesssim 4.2\times 10^{-4}.
\label{bsgb} \eeq
The computation of $\bsg$ in the {\tt micrOMEGAs} package
presented in \cref{microbsg} includes \cite{nlobsg}
\emph{next-to-leading order} (NLO) QCD corrections to the charged
Higgs boson ($H^\pm$) contribution, the $\tan\beta$ enhanced
contributions, as well as resummed NLO SUSY QCD corrections. The 
$H^\pm$
contribution interferes constructively with the SM contribution,
whereas the SUSY contribution interferes destructively with the
other two contributions for $\mu>0$. The SM plus the $H^\pm$ and
SUSY contributions initially increases with $\mx$ and yields a
lower bound on $\mx$ from the lower bound in Eq.~(\ref{bsgb}).
(For higher values of $\mx$, it starts mildly decreasing.)

\paragraph{The Branching Ratio $\bmm$ of
$B_s\to\mu^+\mu^-$.} The rare decay $B_s\to \mu^+\mu^-$ occurs via
$Z$ penguin and box diagrams in the SM and, thus, its branching
ratio is highly suppressed. The SUSY contribution, though,
originating \cite{bsmm, mahmoudi} from neutral Higgs bosons in
chargino-, $H^\pm$-, and $W^\pm$-mediated penguins behaves as
$\tan^6\beta/m^4_A$ ($m_A$ is the mass of the CP-odd Higgs
boson $A$) and hence is particularly important for large
$\tan\beta$'s, especially after the new more stringent $95\%$ c.l. 
upper bound
\beq \bmm\lesssim1.08\times10^{-8} \label{bmmb} \eeq
recently reported by CMS and LHCb \cite{bmmexp1}. This new 
bound significantly reduces the previous bound \cite{bmmexp}, 
which we had adopted in \cref{yqu}. The bound in Eq.~(\ref{bmmb}) 
implies a lower bound on $\mx$ since $\bmm$ decreases as 
$m_{\rm LSP}$ increases.

\paragraph{The Branching Ratio
${\rm BR}\lf B_u\to \tau\nu\rg$ of $B_u\to \tau\nu$.} The purely
leptonic decay $B_u\to \tau\nu$ proceeds via $W^\pm$- and
$H^\pm$-mediated annihilation processes. The SUSY contribution,
contrary to the SM one, is not helicity suppressed and depends on
the mass $m_{H^\pm}$ of the charged Higgs boson since it behaves
\cite{mahmoudi,Btn} as $\tan^4\beta/m^4_{H^\pm}$. The ratio
$\btn$ of the CMSSM to the SM branching ratio of $B_u\to \tau\nu$
increases with $\mx$ and approaches unity. It is to be consistent
with the following $95\%$ c.l. range \cite{bsgexp}:
\beq 0.52\lesssim\btn\lesssim2.04\ .\label{btnb} \eeq
A lower bound on $\mx$ can be derived from the lower bound in this
inequality.

\paragraph{Muon Anomalous Magnetic Moment.}
\label{phenoc} The quantity $\delta a_\mu$, which is defined in
\Sref{sec:intro}, can be attributed to SUSY contributions arising
from chargino-sneutrino and neutralino-smuon loops. The relevant
calculation is based on the formulas of Ref.~\cite{gmuon}. The
absolute value of the result decreases as $\mx$ increases
and its sign is positive for $\mu>0$. On the other hand, the
calculation of $a^{\rm SM}_\mu$ is not yet completely stabilized 
mainly because of the ambiguities in the calculation of the 
hadronic vacuum-polarization contribution. According to the 
evaluation of this contribution in Ref.~\cite{g2davier}, there 
is still a discrepancy between the findings based on the
$e^+e^-$-annihilation data and the ones based on the $\tau$-decay
data -- however, in \cref{Jen}, it was claimed that this 
discrepancy can be considerably ameliorated. Taking into account 
the more reliable calculation based on the $e^+e^-$ data and the
experimental measurements \cite{g2exp} of $a_\mu$, we obtain the
following $95\%$ c.l. range -- cf. \cref{Hagiwara}:
\beq~12.7\times 10^{-10}\lesssim \delta a_\mu\lesssim 44.7\times
10^{-10}. \label{g2b}\eeq
A lower [upper] bound on $\mx$ can be derived from the upper
[lower] bound in \Eref{g2b}. As it turns out, only the upper 
bound on $\mx$ is relevant in our case. Taking into account the
aforementioned computational instabilities and the common 
practice \cite{lhc}, we consider this bound only as an optional 
constraint.

\paragraph{Collider Bounds.} For our analysis, the only relevant
collider bound is the $95\%$ c.l. LEP bound \cite{lepmh} on the
lightest CP-even neutral Higgs boson mass
\beq m_h\gtrsim114.4~{\rm GeV},\label{mhb} \eeq
which gives a lower bound on $\mx$. However, we should keep in 
mind that recent data from ATLAS and CMS \cite{Hlhc} provide a 
$99\%$ c.l. upper bound $m_h\lesssim128~\GeV$ and a hint in 
favor of the range $(125\pm1)~\GeV$. Allowing for a theoretical 
error of $\pm1.5~\GeV$ and adding in quadrature the 
experimental and theoretical uncertainties, we construct the 
1$\sigma$ range of interest \cite{news1}:
\beq 123.2\lesssim m_h/\GeV\lesssim126.8.\label{mh11} \eeq
The calculation of $m_h$ in the package {\tt SOFTSUSY}
\cite{Softsusy} includes the full one-loop SUSY corrections and
some zero-momentum two-loop corrections \cite{2loops}. The results
are well tested \cite{comparisons} against other spectrum
calculators.

\section{Restrictions on the SUSY Parameters} \label{results}

\begin{figure}[t]\vspace*{-.32cm}
\begin{minipage}{75mm}
\includegraphics[height=4.5in,angle=-90]{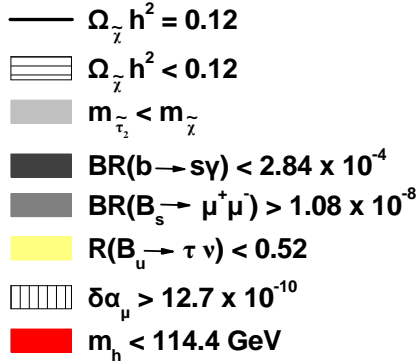}
\end{minipage}
\hfil \hspace*{3mm}\begin{minipage}{70mm}
\vspace*{-.35cm}\caption{\label{Capgx} Summary of the conventions
adopted in Figs.~\ref{Mmx} and \ref{AMgx} for the various
restrictions on the model parameters.}
\end{minipage}\vspace*{-2.cm}
\end{figure}

\begin{figure}[!tb]\vspace*{-0.55cm}
\centering
\includegraphics[width=55mm,angle=-90]{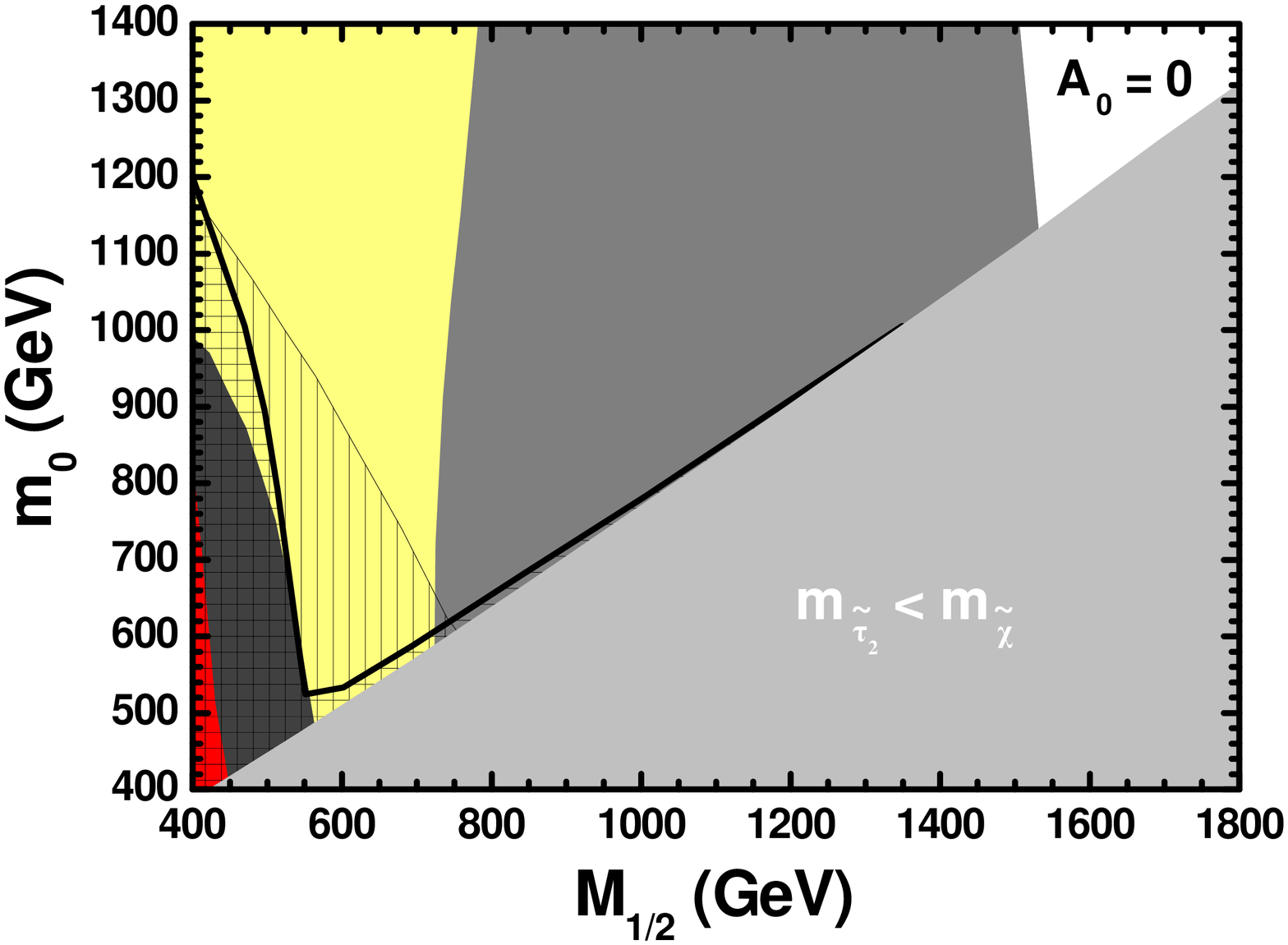}
\includegraphics[width=55mm,angle=-90]{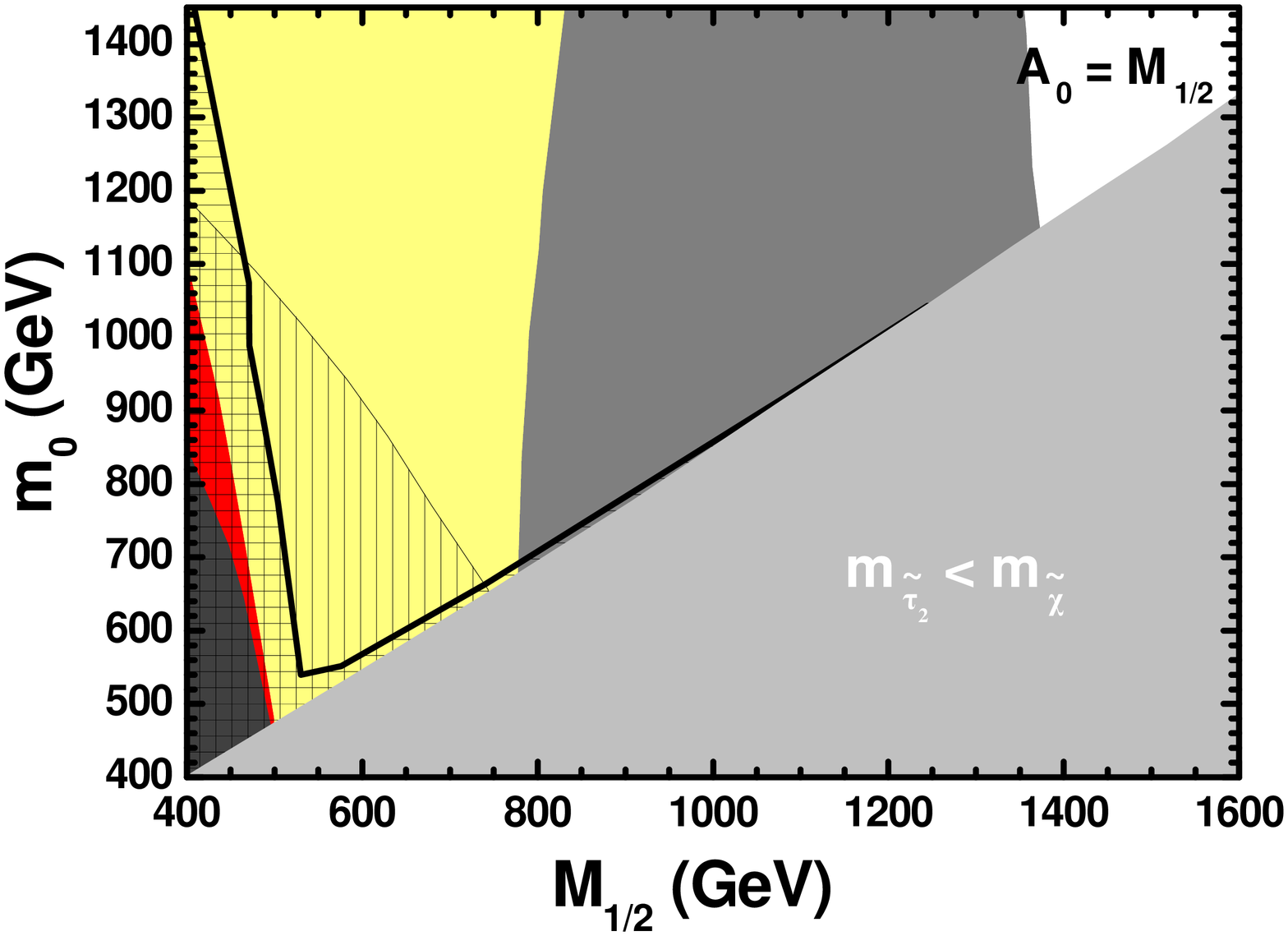}\\
\includegraphics[width=55mm,angle=-90]{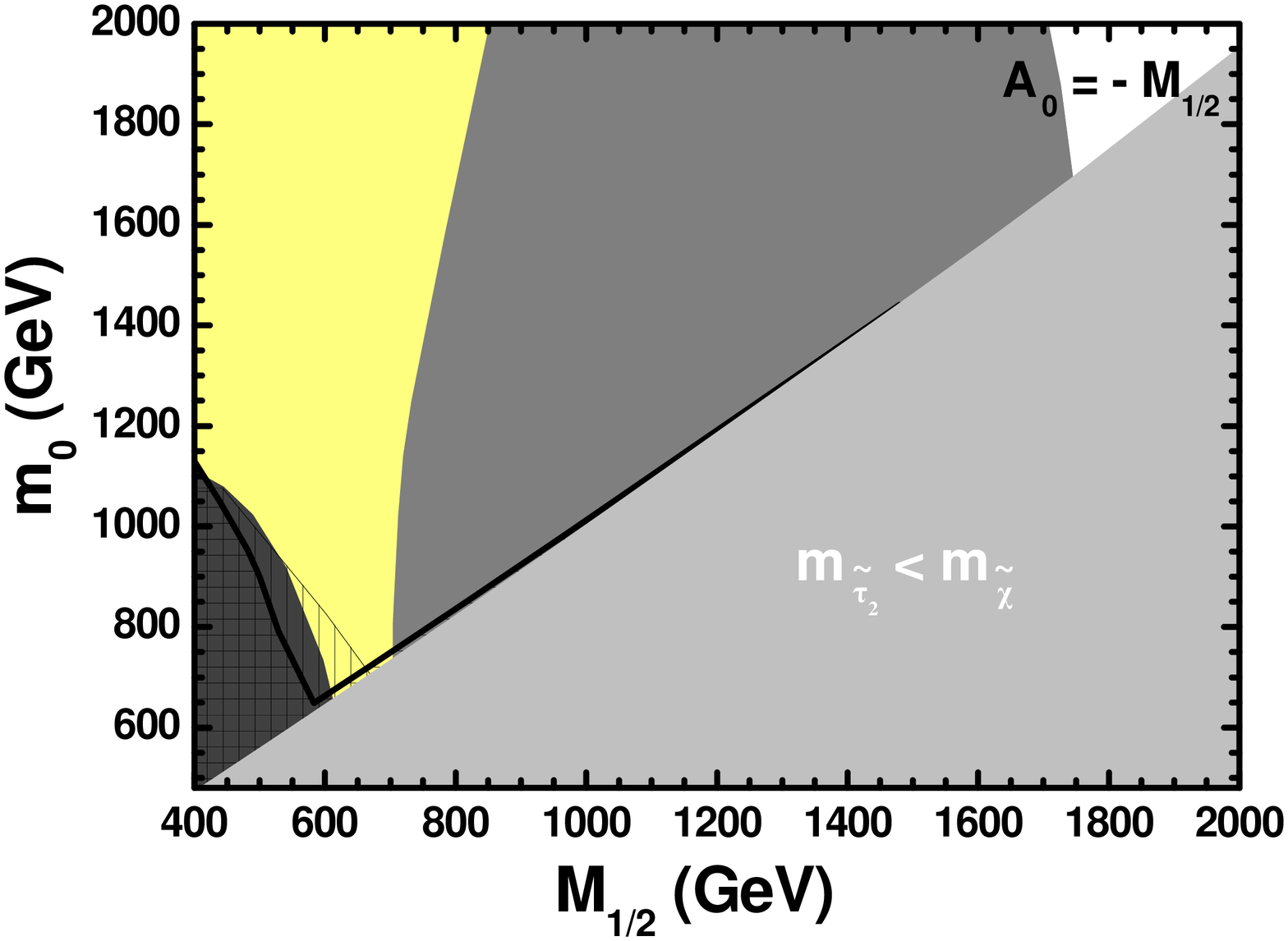}
\includegraphics[width=55mm,angle=-90]{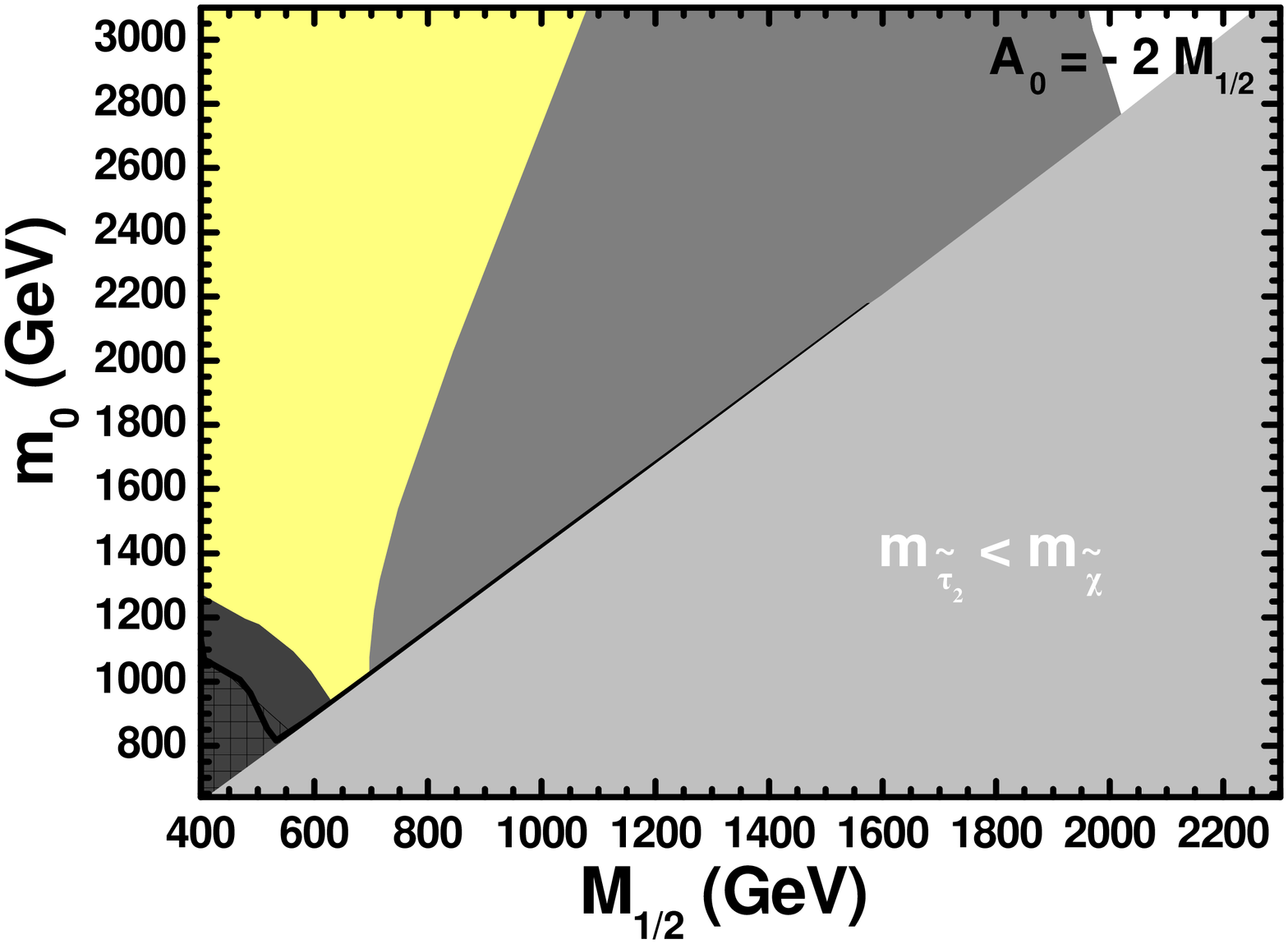}
\caption{The restrictions in the $\Mg-m_{0}$ plane for various values
of $A_0/\Mg$ indicated in the graphs. The conventions adopted are
described in \Fref{Capgx}.} \label{Mmx}
\end{figure}

\begin{figure}[!tb]\vspace*{-0.55cm}
\begin{minipage}{85mm}
\includegraphics[width=65mm,angle=-90]{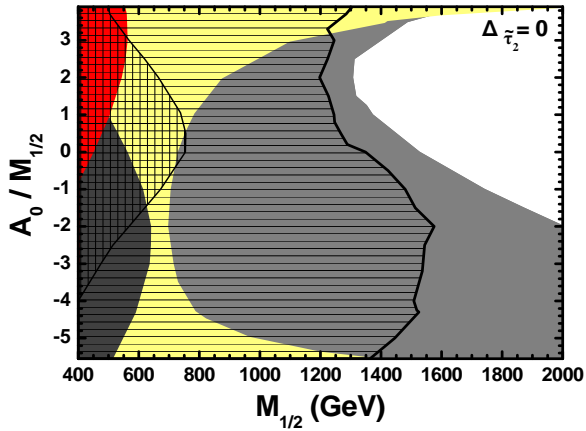}
\end{minipage}
\hfil \hspace*{2mm}\begin{minipage}{60mm}
\vspace*{2.5cm}\caption{The restrictions in the
$M_{1/2}-A_{0}/M_{1/2}$ plane for $\Delta_{\tilde\tau_2}=0$
following the conventions of \Fref{Capgx}.} \label{AMgx}
\end{minipage}
\end{figure}

Imposing the requirements described above, we can delineate the
allowed parameter space of our model. Throughout our investigation, 
we consider the central values for the SM parameters $M_t$,
$m_b(M_Z)$, $m_\tau(M_Z)$, and $\alpha_s(M_Z)$. We adopt the
following conventions for the various lines and regions in the
relevant figures (Figs.~\ref{Mmx} and \ref{AMgx}) -- see
\Fref{Capgx}:

\begin{itemize}

\item On the solid black line, \Eref{cdmb} is saturated.

\item The horizontally hatched region is allowed by \Eref{cdmb}.

\item In the light gray region, the lightest stau $\tilde\tau_2$
is lighter than $\xx$.

\item The dark gray region is excluded by the lower bound in
\Eref{bsgb}.

\item The gray region is excluded by \Eref{bmmb}.

\item The yellow region is excluded by the lower bound in
\Eref{btnb}.

\item The vertically hatched region is favored by the lower bound
in \Eref{g2b}.

\item The red region is excluded by \Eref{mhb}.

\end{itemize}

Note that the upper bounds in Eqs.~(\ref{bsgb}), (\ref{btnb}), 
and (\ref{g2b}) do not restrict the parameters of our model. 
The region with $\tilde\tau_2$ lighter than $\xx$ can not be 
excluded if the LSP is a neutral sparticle other than $\xx$. 
One should, though, make sure that the decay of $\tilde\tau_2$ 
to the LSP does not destroy the predictions of the standard 
BBN \cite{gravitino,axino}.

We present the restrictions from all the requirements imposed in
the $\Mg-m_{0}$ plane for $A_0/\Mg=0$, $1$, $-1$, and $-2$ in
\Fref{Mmx}. From the relevant data, we observe that the lower
bound in \Eref{btnb} is fulfilled for the mass of the CP-odd Higgs
boson $m_A\gtrsim 520~\GeV$ and almost independently of the other
parameters. Note also that, for $\AMg=-1$ and $-2$, the bound in
\Eref{mhb} is violated for $M_{1/2}<400~\GeV$ and, consequently,
does not appear in the relevant diagrams. It is obvious that, for
all the $A_0/\Mg$'s considered in \Fref{Mmx}, we are left with no
region allowed by all the restrictions of \Sref{sec:pheno}. This 
is due to the fact that the constraint in \Eref{cdmb}, which 
necessarily holds if $\tilde\chi$ is the LSP, is nowhere fulfilled 
simultaneously with the bound in \Eref{bmmb}. Note, finally, that 
the upper bound on $\Mg$ from the lower bound in \Eref{g2b} is also 
nowhere satisfied simultaneously with the bound in \Eref{bmmb} for 
the values taken for $\AMg$ in \Fref{Mmx}. However, this conflict 
is less serious since, as we already explained 
in Sec.~\ref{sec:pheno}, the lower bound in \Eref{g2b} is considered 
here only as an optional constraint.   

The constraint in \Eref{cdmb} is, in general, satisfied in two
well-defined distinct regions in the diagrams of \Fref{Mmx}, 
which are:

\begin{itemize}

\item The region to the left of the almost vertical part of the
line corresponding to the upper bound on $\Mg$ from \Eref{cdmb},
where the neutralino annihilation via the $s$-channel exchange 
of a CP-odd Higgs boson $A$ is by far the dominant 
(co)annihilation process. However, this region is excluded by 
the constraints in Eqs.~(\ref{bmmb}) and (\ref{btnb}). On the 
other hand, it is well known -- see e.g. Refs.~\cite{cmssm2, qcdm} 
-- that this region is extremely sensitive to variations of 
$m_b(M_Z)$. Indeed, we find that, as $m_b(M_Z)$ decreases, 
the $A$-boson mass $m_A$ increases and approaches $2m_{\tilde\chi}$. 
The $A$-pole neutralino annihilation is then enhanced and 
$\Omega_{\tilde\chi}h^2$ is drastically reduced causing an 
increase of the upper bound on $\Mg$. However, even if we reduce 
$m_b(M_Z)$, we do not find any $A$-pole neutralino annihilation 
region which is allowed by the requirements in Eqs.~(\ref{bmmb}) 
and (\ref{btnb}).

\item The narrow region which lies just above the light gray area
with $\tilde\tau_2$ lighter than the neutralino, where bino-stau 
coannihilations
\cite{cmssm1,cdm} take over leading to a very pronounced reduction
of $\Omx$. A large portion of this region survives after the
application of all the other requirements of \Sref{sec:pheno}
except for that in Eq.~(\ref{bmmb}). To get a better
understanding of this region, we can replace the parameter $m_0$ by
the relative mass splitting 
$\Dst=(m_{\tilde\tau_2}-m_{\tilde\chi})/m_{\tilde\chi}$ between 
$\tilde\chi$ and the lightest stau, which controls the strength 
of bino-stau coannihilations. The coannihilation region then 
approximately corresponds to $\Dst=0-0.25$. It is evident from
\Fref{Mmx} that the slope of the boundary line with $\Dst=0$
increases as $\AMg$ moves away from zero in both directions. Note
that this slope in our model turns out to be larger than the one
obtained in other versions of the CMSSM -- cf. \cref{cmssm1} --
with lower values of $\tan\beta$. As a consequence, small
variations of $m_0$ or $\Mg$ lead, in our model, to more drastic
variations in $\Dst$.

\end{itemize}

To investigate further whether the incompatibility between the
constraints in \Erefs{cdmb}{bmmb} extends to all possible
$\AMg$'s, we focus on the coannihilation regime and construct the
regions allowed by all the restrictions of \Sref{sec:pheno} in the
$\Mg-A_0/\Mg$ plane for $\Dst=0$. We depict our results in
\Fref{AMgx}. The choice $\Dst=0$ ensures the maximal possible
reduction of $\Omx$ due to the $\tilde\chi-\tilde\tau_2$
coannihilation. So, for a given value of $A_0/\Mg$, the maximal 
$\Mg$ or $\mx$ allowed by \Eref{cdmb}, which holds under the 
assumption that the neutralino is the LSP, corresponds to 
$\Dst=0$. We find that, for $A_0/\Mg<0$, processes with
$\tilde\tau_2\tilde\tau_2^\ast$ in the initial state and $W^\pm
W^\mp$, $W^\pm H^\mp$ in the final one become more efficient (with
a total contribution to the effective cross section of about $14$
to $22\%$ as $A_0/\Mg$ decreases from $0$ to $-2$) and so
coannihilation is strengthened and $\mx$'s larger than in
the $A_0/\Mg>0$ case are allowed by \Eref{cdmb}. The overall
maximal $\Mg\simeq1575~\GeV$ or $\mx\simeq722~\GeV$ allowed by 
\Eref{cdmb} is encountered at $A_0/\Mg\simeq-2$ yielding 
$\bmm=1.82\times10^{-8}$. Comparing the above upper bound on $\Mg$
with the corresponding one in Fig.~3 of \cref{yqu} (represented by 
a solid and two dotted black lines), we observe that here the bound 
is considerably enhanced in the region of low as well as the region 
of large values of $\AMg$ since we do not consider the constraint 
from the lower bound on $\delta a_\mu$ from the $\tau$-based 
calculation. However, it always remains smaller than the lower 
bound on $\Mg$ derived from Eq.~(\ref{bmmb}) -- note 
that for $3.7\lesssim\AMg\lesssim3.9$ the overall lower bound on 
$\Mg$ is derived from Eq.~(\ref{btnb}). Indeed, the smallest lower 
bound on $\Mg=1306~\GeV$ or $\mx\simeq590~\GeV$  is found at 
$A_0/\Mg\simeq2$ yielding $\Omx=0.15$. Note that increasing $\Dst$ 
within the range $0-0.25$ does not alter the boundaries of the 
various constraints in any essential way, except the solid line 
which is displaced to the left shrinking, thereby, the area allowed 
by \Eref{cdmb} considerably. Needless to say that the more 
stringent optional upper bound on $\Mg$ from the lower bound in 
\Eref{g2b} is also not compatible with the constraint in 
\Eref{bmmb}. Consequently, for every $A_0$, there is no range 
of parameters simultaneously allowed by all the constraints 
and, therefore, $\xx$ can be now excluded as a CDM particle in our 
model.

\begin{figure}[t]\vspace*{-0.6cm}
\begin{minipage}{75mm}
\includegraphics[height=9cm,angle=-90]{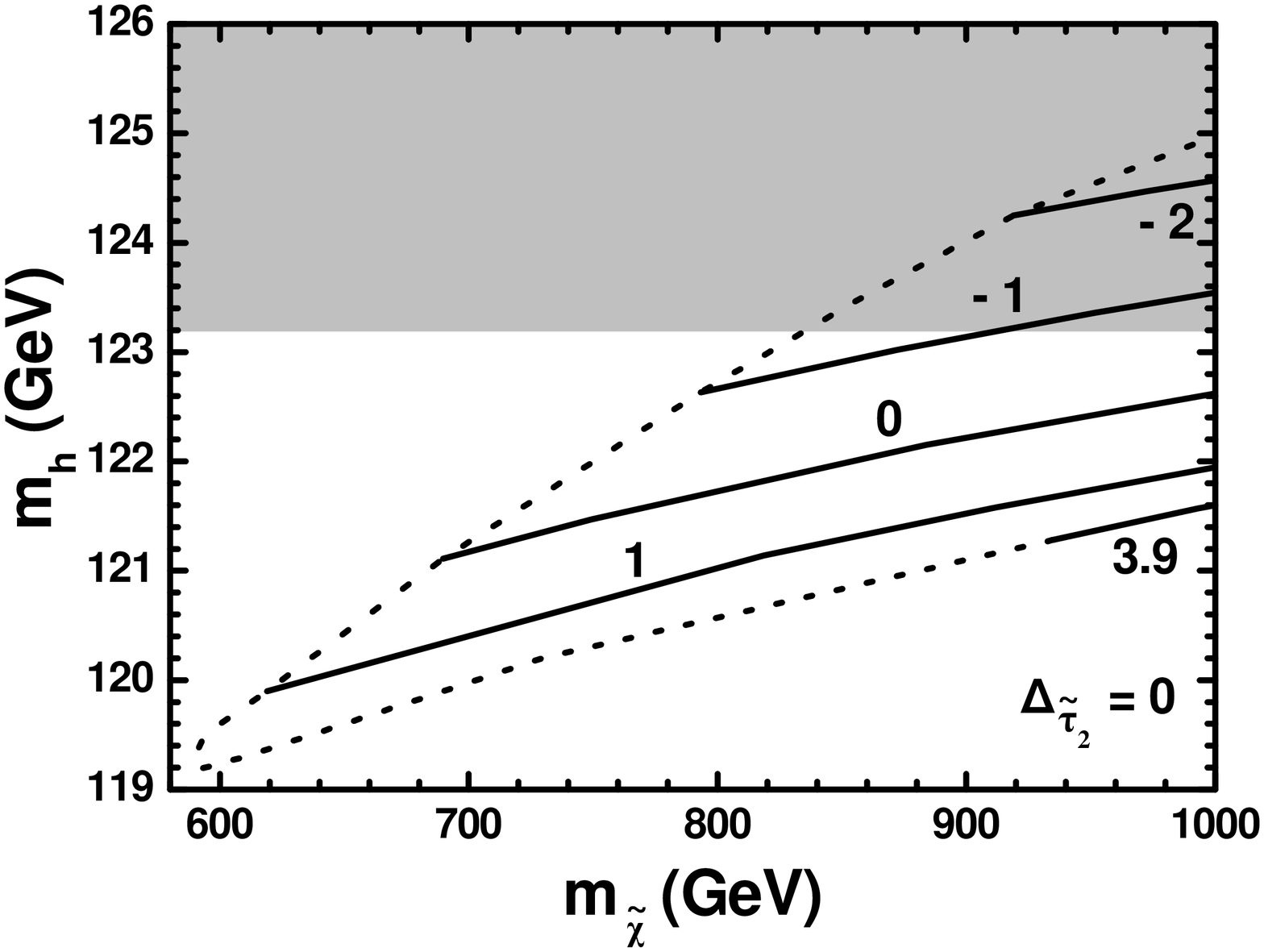}
\end{minipage}
\hfil
\begin{minipage}{65mm}
\begin{center}
\lineup
\begin{tabular}{@{}ccc@{}}
\br
$\AMg$&$m_{\tilde\chi}|_{\rm min}$&{$\Omega_{\tilde\chi} h^2$}\cr\mr
$-2$&$919~\GeV$&{$0.22-0.32$}\cr
$-1$&$793~\GeV$&{$0.17-0.31$}\cr
$0$&$688~\GeV$&{$0.15-0.38$}\cr
$1$&$619~\GeV$&{$0.16-0.44$}\cr
$3.9$&$935~\GeV$&{$0.31-0.34$}\cr \br
\end{tabular}
\end{center}
\end{minipage}
\caption{The variation of $m_h$ as a function of  $\mx$ for
$\Dst\simeq0$ and various $A_0/M_{1/2}$'s indicated on the
curves and in the table included in this figure. In this table, 
listed are also the minimal $\mx$'s for which the inequality in 
\Eref{bmmb} and the lower bound in \Eref{btnb} are satisfied 
and the corresponding ranges of $\Omx$ as $\mx$ increases from 
$m_{\tilde\chi}|_{\rm min}$ to about $1~{\rm TeV}$. The minimal 
$m_h$ for all the values of 
$A_0/M_{1/2}$ is also depicted by a dotted line. The part of the 
region of \Eref{mh11} preferred by the recent LHC data which 
lies in the panel is painted light gray.} \label{mxh}
\end{figure}

The exclusion, in our model, of $\xx$ as a CDM candidate, 
resulting from the incompatibility between \Erefs{cdmb}{bmmb}, is 
further strengthened if one tries to reconcile \Erefs{cdmb}{mh11}. 
Indeed, the tension between the neutralino CDM and the new data on 
$m_h$ is quite generic within the CMSSM since the fulfillment of 
\Eref{mh11} requires a very heavy SUSY spectrum, which leads to 
conflict with \Eref{cdmb} -- cf. Refs.~\cite{news1,news2}. It 
would be interesting to investigate this issue in our model, which 
yields large values of $\tan\beta$ and so \Eref{mh11} can be 
possibly satisfied with a lighter SUSY spectrum relative to other 
versions of the CMSSM with lower $\tan\beta$'s. Our results are 
presented in \Fref{mxh}, where we draw $m_h$ (bold lines) versus 
$\mx$ for $\Delta_{\tilde\tau_2}\simeq 0$ and $\AMg=-2$, $-1$, $0$, $1$, 
and $3.9$. In the table included in \Fref{mxh}, we also list the 
minimal $\mx$'s, $m_{\tilde\chi}|_{\rm min}$, for which the 
inequality in \Eref{bmmb} and the lower bound in \Eref{btnb} are 
satisfied for given values of $\AMg$ as well as the corresponding 
ranges of $\Omega_{\tilde\chi} h^2$ as $\mx$ varies from 
$m_{\tilde\chi}|_{\rm min}$ to about $1~{\rm TeV}$. Since we take 
$\Dst\simeq0$, the derived $\Omx$ takes its minimal possible value. 
In this plot, we also depict by a dotted line the value of 
$m_{\tilde\chi}|_{\rm min}$ for all possible $\AMg$'s and 
$\Dst\simeq0$. This line terminates at $\AMg\simeq3.9$ since beyond 
this value the stability of the electroweak vacuum fails. The overall 
minimal $m_h\simeq119~\GeV$ is encountered at $\AMg\simeq2$ and 
$\mx\simeq590~\GeV$. It is interesting to note that $m_h$ increases 
with $\mx$ (or $M_{1/2}$) and as $A_0$ decreases and, eventually, 
this mass enters inside 
the gray region in \Fref{mxh}, which is preferred, at 1$\sigma$, by 
the recent LHC searches. However, it is obvious from Fig.~\ref{mxh}
and the values of $\Omx$ in the table included in this figure that
\Erefs{cdmb}{mh11} are incompatible even in our model. 

Departure from $\Dst=0$ is not expected to alter drastically our 
predictions as regards the value of $m_h$ since $m_h$ depends 
crucially on $M_{1/2}$, but only mildly on $m_0$. Moreover, one 
can deduce from the slop of the left boundary of the gray regions 
in \Fref{Mmx} that, increasing $\Dst$, smaller $M_{1/2}$'s 
and, therefore, slightly lighter Higgs masses are permitted by 
\Eref{bmmb}. Note that, since $\xx$ cannot be the LSP, both signs 
of $\Dst$ are possible. In particular $\Dst>0$ [$\Dst<0$] 
corresponds to $\xx$ [$\tilde\tau_2$] being the LOSP. The critical 
case $\Dst\simeq0$ gives the minimal possible relic abundance of 
the LOSP in both cases due to the coannihilation effect and, 
therefore, the maximal possible mass of the LSP if this is produced 
mainly non-thermally -- see Refs.~\cite{gravitino,axino,Baerax}.

Having in mind mostly the latter possibility, we proceed in the
presentation of our predictions for the sparticle and the Higgs
boson spectrum of our model, which may be observable at the LHC.
In \Tref{spectrum}, we list the model input and output parameters,
the masses in $\GeV$ of the sparticles -- neutralinos
$\tilde\chi$, $\tilde\chi_2^{0}$, $\tilde{\chi}_{3}^{0}$,
$\tilde{\chi}_{4}^{0}$, charginos $\tilde{\chi}_{1}^{\pm}$,
$\tilde{\chi}_{2}^{\pm}$, gluinos $\tilde{g}$, squarks
$\tilde{t}_1$, $\tilde{t}_2$, $\tilde{b}_1$, $\tilde{b}_2$,
$\tilde{u}_{L}$, $\tilde{u}_{R}$, $\tilde{d}_{L}$,
$\tilde{d}_{R}$, and sleptons $\tilde\tau_1$, $\tilde\tau_2$,
$\tilde\nu_\tau$, $\tilde{e}_L$, $\tilde{e}_R$, $\tilde{\nu}_{e}$
-- and the Higgs bosons ($h$, $H$, $H^\pm$, $A$), and the values
of the various low energy observables for $\Dst\simeq 0$, $A_0/\Mg=-3$,
$-2$, $-1$, $0$, and $2$ and for the minimal $\Mg$ allowed 
by Eq.~(\ref{bmmb}) in each case. We  consider the squarks and 
sleptons of the two first
generations as degenerate. From the values of the various
observable quantities, we see that the bound in \Eref{cdmb} and 
the optional lower bound in \Eref{g2b} are violated. So, the 
lightest neutralino cannot be the LSP. It is also very interesting 
to observe that the predicted values of $m_h$ lie close or even 
inside the range in \Eref{mh11} favored by \cref{Hlhc} -- cf. 
Refs.~\cite{news1,news2}.

The deviation from YU can be estimated by defining \cite{nova} the
relative splittings $\delta h_{b}$ and $\delta h_{\tau}$ at
$M_{\rm GUT}$ through the relations:
\beq \delta h_{b}\equiv\frac{h_{b}-h_t}{h_t}=-\frac{2c}{1+c}=
-\delta h_{\tau}\equiv\frac{h_t-h_{\tau}}{h_t}\cdot\eeq
Along the dotted line of \Fref{mxh}, the ranges of the parameters
$c,~\delta h_{\tau},~\delta h_{b}$, and $\tan\beta$ are
\bea \nonumber 0.148\lesssim c\lesssim0.16,\>\>\>
0.26\lesssim\delta h_{\tau}=-\delta h_b \lesssim0.28,\>\>\>
56.2\lesssim \tan\beta\lesssim 56.9. \eea
Let us underline that, although the required deviation from YU is
not so small, the restrictions from YU are not completely lost
since $\tan\beta$ remains large -- close to 60 -- and that the
deviation from exact YU is generated within well-motivated SUSY 
GUTs described in \cref{qcdm}.

It is worth emphasizing that our results do not invalidate the
$\xx$ candidacy for a CDM particle in all versions of the CMSSM 
with Yukawa quasi-unification. This is because the (monoparametric) 
condition of \Eref{minimal}, which we considered here, is only a 
simplified case of the Yukawa quasi-unification conditions shown 
in Eq.~(15) of \cref{qcdm}, which depend on one real and two 
complex parameters. Actually, the investigation of the viability 
of $\xx$ as a CDM candidate within the CMSSM with Yukawa 
quasi-unification conditions more complicated than the one in 
\Eref{minimal} derived from the GUT models of \cref{qcdm} is 
under consideration. Alternatively, our present model may be 
perfectly consistent with data if we avoid the restriction from 
\Eref{cdmb} by assuming that the LSP is the axino 
\cite{axino, Baerax} with mass a little lower than $\mx$ and 
that the reheat temperature is adequately low.

\clearpage
\begin{table}[!ht]
\caption{Input and output parameters, masses of the sparticles and
Higgs bosons, and values of the low energy observables for 
$\Dst\simeq0$, five values of $A_0/\Mg$, and the minimal $M_{1/2}$.} 
\begin{center}
\lineup{\small
\begin{tabular}{c@{\hspace{0.1cm}}c@{\hspace{0.3cm}}c@{\hspace{0.3cm}}c@{\hspace
{0.3cm}}c@{\hspace{0.3cm}}c}
\br \multicolumn{6}{c}{Input parameters}\\\mr
$A_0/\Mg$ &$-3$ &$-2$ &$-1$ &$0$ &$2$  \\
$c$  &$0.1589$ &$0.1592$ &$0.1585$ &$0.153$&$0.1475$  \\
$\Mg/\GeV$ &$2355.02$ & $2019.6$ &$1744.65$ &$1530.3$&$1317.65$  \\
$m_0/\GeV$ &$4292.06$ &$2760.01$ &$1691.01$ &$1132.04$&$1542.88$ \\ \mr
\multicolumn{6}{c}{Output parameters}\\\mr
$\tan\beta$ &$57.1$ & $56.9$ &$56.3$ & $56.2$ & $56.2$  \\
$100\delta h_\tau(M_{\rm GUT})$ &$27.4$ & $27.4$ & $27.3$ & $26.5$ & $25.7$  \\
$\mu/\GeV$ &$3704.64$ & $2755$ & $2059$ & $1588$ & $1250$  \\ \mr
\multicolumn{6}{c}{Masses in ${\rm GeV}$ of sparticles and
Higgs bosons}\\\mr
$\tilde\chi$&$1089.2$ & $926.5$ &$794.2$ &$692.5$ &$595.2$  \\
$\tilde\chi_2^{0}$ &$2087.2$ &$1777.7$ &$1524.2$ &$1325.4$ &$1130.1$  \\
$\tilde{\chi}_{3}^{0}$ &$3688.5$ &$2747.6$ &$2057.3$ &$1588.9$ &$1254.8$  \\
$\tilde{\chi}_{4}^{0}$ &$3789.7$ &$2750.3$ &$2062.7$ &$1600.5$ &$1279.9$  \\
$\tilde{\chi}_{1}^{\pm}$ &$3690.1$ & $2750.6$ &$2062.8$ &$1600.4$ &$1279.6$  \\
$\tilde{\chi}_{2}^{\pm}$ &$2087.3$ &$1777.8$ &$1524.3$ &$1325.5$&$1130.2$  \\
$\tilde{g}$&$5190.9$ &$4454$ &$5190.9$ &$3388.6$&$2981.2$ \\ \mr
%
$\tilde{t}_1$ &$4336.3$ &$3608.2$ &$3094.3$ &$2752.2$ &$2567.8$ \\
$\tilde{t}_2$ &$3593.2$ &$3084.5$ &$2709.5$ &$2449.2$&$2303.1$ \\
$\tilde{b}_1$ &$4514.1$ &$3653.4$ &$3097.4$ &$2747.9$&$2575.7$ \\
$\tilde{b}_2$ &$4310.6$ &$3561.2$ &$3005.7$ &$2644.1$ &$2508.8$ \\
$\tilde{u}_{L}$ &$6215.9$ &$4786$ &$3815.1$ &$3231.7$ &$3051.1$ \\
$\tilde{u}_{R}$ &$6047.6$ &$4624.8$ &$3661.4$ &$3090.5$ &$2941.3$  \\
$\tilde{d}_{L}$ &$6216.3$& $4786.5$ & $3815.8$ &$3232.5$ &$3052$  \\
$\tilde{d}_{R}$ &$6026.1$ &$4604.4$ &$3641.9$ &$3072.7$&$2927.7$ \\\mr
$\tilde\tau_1$ &$3447.3$ & $2413.2$ &$1721.0$ &$1354.4$&$1436.4$ \\
$\tilde\tau_2$ &$1089.9$ &$927.1$ &$794.4$ &$692.5$&$595.7$ \\
$\tilde\nu_\tau$ &$3443.9$ &$2407.7$ &$1712.4$ &$1343.3$&$1430.2$ \\
$\tilde{e}_L$ &$4582.6$ &$3085.1$ &$2070.4$ &$1544.9$&$1789.6$ \\
$\tilde{e}_R$ &$4389.7$ &$2869.5$ &$1821.4$&$1278.5$&$1626.4$ \\
$\tilde{\nu}_{e}$ &$4581.6$ &$3083.7$ &$2068.6$ &$1542.6$ &$1787.5$ \\\mr
%
$h$  &$126.17$ &$124.3$ &$122.68$ &$121.15$ &$119.30$ \\
$H$ &$1463.72$ &$1334.6$ &$1181.53$ &$1012.4$ &$730.78$  \\
$H^{\pm}$  &$1466.38$ &$1337.6$ &$1185.04$ &$1016.55$ &$736.91$ \\
$A$ &$1463.99$ &$1334.9$ &$1182.00$ & $1013$&$732$  \\\mr
\multicolumn{6}{c}{Low energy observables}\\\mr
%
$10^4\bsg$ &$3.25$ &$3.23$ &$3.22$ &$3.23$ &$3.35$ \\
$10^8\bmm$  &$1.08$ &$1.08$ &$1.08$ &$1.08$ &$1.08$ \\
$\btn$  &$0.929$ &$0.915$ &$0.893$ &$0.856$ &$0.736$ \\
$10^{10}\Dam$ &$0.565$ &$1.09$ &$2.04$ &$3.27$ &$3.4$  \\\mr
$\Omx$ &$0.301$ &$0.219$ &$0.167$ &$0.152$&$0.151$ \\
\br
\end{tabular}}
\end{center}
\label{spectrum}
\end{table}

\section{Conclusions} \label{con}

We performed a revised scan of the parameter space of the CMSSM
with $\mu>0$ applying a suitable Yukawa quasi-unification
condition predicted by the SUSY GUT model of Ref.~\cite{qcdm},
which has been constructed in order to remedy the $b$-quark mass
problem arising from exact YU and universal
boundary conditions. We took into account updated constraints from
collider and cosmological data. These constraints originate from
the CDM abundance in the universe, $B$ physics ($b \rightarrow
s\gamma$, $B_s\to \mu^+\mu^-$, and $B_u\to\tau\nu$),
$\delta\alpha_\mu$, and $m_h$. Although the neutralino-stau
coannihilations drastically reduce the neutralino relic 
abundance and, thus, enhance the upper bound on $\mx$ implied 
by the assumption that the neutralino is a CDM particle, they 
do not quite succeed to bring it to an acceptable level 
compatible with the lower bound on $\mx$ induced by $\bmm$. 
Therefore, -- contrary to our findings in \cref{yqu} -- $\xx$ 
is excluded as CDM particle by the combination of the 
constraints from $\bmm$ and CDM.  As a consequence, the model 
can become consistent with observations only if the LSP is a 
SUSY particle other than the neutralino. This could be the 
axino or the gravitino and can account for the present CDM 
abundance in the universe. It is interesting to note that, in 
this case, the lowest predicted $m_h$ is enhanced and gets 
closer to the range favored by the recent preliminarily 
results announced by LHC. 

\ack We would like to thank A.~Djouadi, N.~Mahmoudi, K.A.~Olive,  
P.~Paradisi, Q.~Shafi, and J.~Wells for useful discussions. This 
work was supported by the European Union under the Marie Curie 
Initial Training Network `UNILHC' PITN-GA-2009-237920 and the 
Greek Ministry of Education, Lifelong Learning and Religious 
Affairs and the Operational Program: Education and Lifelong 
Learning `HERACLITOS II'.

\section*{References}

\def\ijmp#1#2#3{{Int. Jour. Mod. Phys.}
{\bf #1},~#3~(#2)}
\def\plb#1#2#3{{\it Phys. Lett. B }{\bf #1},~#3~(#2)}
\def\zpc#1#2#3{{\it Z. Phys. C }{\bf #1},~#3~(#2)}
\def\prl#1#2#3{{\it Phys. Rev. Lett.}
{\bf #1},~#3~(#2)}
\def\rmp#1#2#3{{\it Rev. Mod. Phys.}
{\bf #1},~#3~(#2)}
\def\prep#1#2#3{{\it Phys. Rep. }{\bf #1},~#3~(#2)}
\def\prd#1#2#3{{\it Phys. Rev. D }{\bf #1},~#3~(#2)}
\def\npb#1#2#3{{\it  Nucl. Phys. }{\bf B#1},~#3~(#2)}
\def\npps#1#2#3{{\it  Nucl. Phys. B (Proc. Sup.)}
{\bf #1},~#3~(#2)}
\def\mpl#1#2#3{{\it  Mod. Phys. Lett.}
{\bf #1},~#3~(#2)}
\def\arnps#1#2#3{{\it  Annu. Rev. Nucl. Part. Sci.}
{\bf #1},~#3~(#2)}
\def\sjnp#1#2#3{{\it  Sov. J. Nucl. Phys.}
{\bf #1},~#3~(#2)}
\def\jetp#1#2#3{{\it  JETP Lett. }{\bf #1},~#3~(#2)}
\def\app#1#2#3{{Acta Phys. Polon.}
{\bf #1},~#3~(#2)}
\def\rnc#1#2#3{{Riv. Nuovo Cim.}
{\bf #1},~#3~(#2)}
\def\ap#1#2#3{{Ann. Phys. }{\bf #1},~#3~(#2)}
\def\ptp#1#2#3{{Prog. Theor. Phys.}
{\bf #1},~#3~(#2)}
\def\apjl#1#2#3{{Astrophys. J. Lett.}
{\bf #1},~#3~(#2)}
\def\apjs#1#2#3{{Astrophys. J. Suppl.}
{\bf #1},~#3~(#2)}
\def\n#1#2#3{{Nature }{\bf #1},~#3~(#2)}
\def\apj#1#2#3{{Astrophys. J.}
{\bf #1},~#3~(#2)}
\def\anj#1#2#3{{Astron. J. }{\bf #1},~#3~(#2)}
\def\mnras#1#2#3{{MNRAS }{\bf #1},~#3~(#2)}
\def\grg#1#2#3{{Gen. Rel. Grav.}
{\bf #1},~#3~(#2)}
\def\s#1#2#3{{Science }{\bf #1},~#3~(#2)}
\def\baas#1#2#3{{Bull. Am. Astron. Soc.}
{\bf #1},~#3~(#2)}
\def\ibid#1#2#3{{\it ibid. }{\bf #1},~#3~(#2)}
\def\cpc#1#2#3{{\it Comput. Phys. Commun.}
{\bf #1},~#3~(#2)}
\def\astp#1#2#3{{\it Astropart. Phys.}
{\bf #1},~#3~(#2)}
\def\epjc#1#2#3{{\it  Eur. Phys. J. C}
{\bf #1},~#3~(#2)}
\def\nima#1#2#3{{\it  Nucl. Instrum. Meth. A}
{\bf #1},~#3~(#2)}
\def\jhep#1#2#3{{\it J. High Energy Phys.}
{\bf #1},~#3~(#2)}
\def\jcap#1#2#3{{\it J. Cosmol. Astropart. Phys.}
{\bf #1},~#3~(#2)}


\begin{thebibliography}{99}

\bibitem{Cmssm0} A.H.~Chamseddine, R.L.~Arnowitt, and P. Nath, 
\prl{49}{1982}{970}; 
P.~Nath, R.L.~Arnowitt, and A.H.~Chamseddine, \npb{227}{1983}{121};  
L.J.~Hall, J.D.~Lykken, and S.~Weinberg, \prd{27}{1983}{2359}.

\bibitem{Cmssm} R.~Arnowitt and P.~Nath, {\it Phys. Rev. Lett.} 
{\bf 69}, 725 (1992);
G.G.~Ross and R.G.~Roberts, {\it Nucl. Phys.} {\bf B377}, 571 (1992);
V.D.~Barger, M.S.~Berger, and P.~Ohmann, 
{\it Phys. Rev.  D} {\bf 49}, 4908 (1994); 
G.L.~Kane, C.F.~Kolda, L.~Roszkowski, and J.D.~Wells,
{\it ibid.} {\bf 49}, 6173 {(1994)}.

\bibitem{cmssm1} J.R.~Ellis, T.~Falk, and K.A.~Olive, 
\plb{444}{1998}{367}; 
J.R.~Ellis, T.~Falk, K.A.~Olive, and M.~Srednicki, 
\astp{13}{2000}{181}; {\bf 15}, 413(E) (2001).

\bibitem{cmssm2} A.B.~Lahanas, D.V.~Nanopoulos, and
V.C.~Spanos, \prd{62}{2000}{023515}; J.R.~Ellis, T.~Falk, 
G.~Ganis, K.A.~Olive, and M.~Srednicki, \plb{510}{2001}{236}.

\bibitem{pana} G.~Lazarides and C.~Panagiotakopoulos, 
\plb{337}{1994}{90}; 
S.~Khalil, G.~Lazarides, and C.~Pallis, \ibid{508}{2001}{327}.

\bibitem{als} B.~Ananthanarayan, G.~Lazarides, and Q.~Shafi, 
\prd{44}{1991}{1613}; \plb{300}{1993}{245}.

\bibitem{leontaris} I.~Antoniadis and G.K.~Leontaris,
\plb{216}{1989}{333}.

\bibitem{jean} R.~Jeannerot, S.~Khalil, G.~Lazarides, and 
Q.~Shafi, \jhep{10}{2000}{012}.

\bibitem{baery}  H.~Baer, M.A.~Diaz, J.~Ferrandis, and X.~Tata,
{\it Phys. Rev. D} {\bf 61}, 111701 (2000); H.~Baer, M.~Brhlik,
M.A.~Diaz, J.~Ferrandis, P.~Mercadante, P.~Quintana, and X.~Tata,
\ibid{63}{2000}{015007}; D.~Auto, H.~Baer, C.~Bal\'azs,
A.~Belyaev, J.~Ferrandis, and X.~Tata, \jhep{06}{2003}{023};
D.~Auto, H.~Baer, A.~Belyaev, and T.~Krupovnickas,
\ibid{10}{2004}{066}.

\bibitem{raby} T.~Bla\v zek, R.~Derm\'\i\v sek, and S.~Raby,
{\it Phys. Rev. Lett. }{\bf 88}, 111804 (2002);
\prd{65}{2002}{115004}; R.~Derm\'\i\v sek, S.~Raby, L.~Roszkowski,
and R. Ruiz de Austri, \jhep{04}{2003}{037}.

\bibitem{copw}
M.S.~Carena, M.~Olechowski, S.~Pokorski, and C.E.M.~Wagner,
\npb{426}{1994}{269}; R.~Hempfling, \prd{49}{1994}{6168}; 
L.J.~Hall, R.~Rattazzi, and U.~Sarid, \ibid{50}{1994}{7048}.

\bibitem{pierce} D.M.~Pierce, J.A.~Bagger, K.T.~Matchev, 
and R.~Zhang, \npb{491}{1997}{3}; 
M.S.~Carena, D.~Garcia, U.~Nierste, and C.E.M.~Wagner, 
\ibid{B577}{2000}{88}.

\bibitem{sugra} P.Z.~Skands {\it et al.}, \jhep{07}{2004}{036}.

\bibitem{shafi} S.F.~King and M.~Oliveira, {\it Phys. Rev. D} {\bf 63},
015010 (2001); I.~Gogoladze, R.~Khalid, and Q.~Shafi,
{\it ibid.} {\bf 79}, 115004 (2009); {\bf 80}, 095016 (2009);
I.~Gogoladze, R.~Khalid, S.~Raza, and Q.~Shafi, 
\jhep{12}{2010}{055}.

\bibitem{nath} U.~Chattopadhyay and P.~Nath,
\prd{65}{2002}{075009}; 
U.~Chattopadhyay, A.~Corsetti, and P.~Nath, 
\ibid{66}{2002}{035003}; 
C.~Pallis, \npb{678}{2004}{398}.

\bibitem{qcdm} M.E.~G\'{o}mez, G.~Lazarides, and C.~Pallis, 
{\it Nucl. Phys.} {\bf B638}, 165 (2002).

\bibitem{axilleas} R.~Jeannerot, S.~Khalil, and G.~Lazarides, 
\jhep{07}{2002}{069}; 
G.~Lazarides and A.~Vamvasakis, {\it Phys.\ Rev.\  D} {\bf 76}, 083507 (2007); 
{\bf 76}, 123514 (2007); G.~Lazarides, I.N.R.~Peddie, and A.~Vamvasakis,
{\it ibid.}~{\bf 78}, 043518 (2008); 
G.~Lazarides, arXiv:1006.3636.

\bibitem{muneg} M.E.~G\'{o}mez, G.~Lazarides, and C.~Pallis,
\prd{67}{2003}{097701}.

\bibitem{nova} G.~Lazarides and C.~Pallis, hep-ph/0404266;
hep-ph/0406081.

\bibitem{g2davier} M.~Davier, A.~Hoecker, B.~Malaescu, and Z.~Zhang,
\epjc{71}{2011}{1515}.

\bibitem{Jen} F.~Jegerlehner and R.~Szafron, 
{\it Eur. Phys. J. C} {\bf 71}, 1632  (2011).

\bibitem{yqu} N.~Karagiannakis, G.~Lazarides, and C.~Pallis,
\plb{704}{2011}{43}.

\bibitem{shafi11} S.~Dar, I.~Gogoladze, Q.~Shafi, and C.S.~Un, 
{\it Phys. Rev. D} {\bf 84}, 085015 (2011).

\bibitem{bmmexp1} CMS and LHCb Collaborations, CMS-PAS-BPH-11-019,
LHCb-CONF-2011-047,\\ 
{\tt http://cdsweb.cern.ch/record/1374913/files/BPH-11-019-pas.pdf.}

\bibitem{Hlhc} Update on the Standard Model Higgs searches in ATLAS 
and CMS, talks by F. Gianotti and G. Tonelli, 13/12/2011, CERN, 
{\tt http://indico.cern.ch/conferenceDisplay.py?confId=164890}.

\bibitem{Softsusy} B.C.~Allanach, 
{\it Computer Physics Commun.} {\bf 143}, 305 (2002).

\bibitem{micro}
G.~Belanger, F.~Boudjema, A.~Pukhov, and A.~Semenov, 
{\tt http://lapth.in2p3.fr/micromegas}; 
G.~Belanger, F.~Boudjema, P.~Brun, A.~Pukhov, S.~Rosier-Lees, 
P.~Salati, and A.~Semenov,
{\it Comput. Phys. Commun.} {\bf 182}, 842 (2011).

\bibitem{lhc} R.~Trotta, F.~Feroz, M.P.~Hobson, L.~Roszkowski, and
R.~Ruiz de Austri, \jhep{12}{2008}{024}; A.~Belyaev, S.~Dar, 
I.~Gogoladze, A.~Mustafayev, and Q.~Shafi, arXiv:0712.1049 [hep-ph];
O.~Buchmueller {\it et al.}, {\it Eur. Phys. J. C} {\bf 64}, 391 (2009);
Y.~Akrami, P.~Scott, J.~Edsjo, J.~Conrad, and L.~Bergstrom,
\jhep{04}{2010}{057}; L.~Roszkowski, R.~Ruiz de Austri, and
R.~Trotta, {\it Phys. Rev. D} {\bf 82}, 055003 (2010); O.~Buchmueller
{\it et al.}, arXiv:1110.3568.

\bibitem{baerlhc} H.~Baer, S.~Kraml, S.~Sekmen, and H.~Summy,
\jhep{03}{2008}{056}; {\bf 10}, 079 (2008); H.~Baer, S.~Kraml, and
S.~Sekmen, \ibid{09}{2009}{005}; H.~Baer, S.~Kraml, A.~Lessa, and
S.~Sekmen, \ibid{02}{2010}{055}; H.~Baer, S.~Kraml, A.~Lessa,
S.~Sekmen, and H.~Summy, {\it Phys. Lett. B} {\bf 685}, 72 (2010).

\bibitem{pdata} K.~Nakamura {\it et al.}~[Particle Data Group],
{\it J. Phys. G} {\bf 37}, 075021 (2010).

\bibitem{baermb} H.~Baer, J.~Ferrandis, K.~Melnikov, and X.~Tata, 
{\it Phys. Rev. D} {\bf 66}, 074007 (2002); 
K.~Tobe and J.D.~Wells, {\it Nucl. Phys.} {\bf B663}, 123 (2003).

\bibitem{mtmt} Tevatron Electroweak Working Group [CDF and D0
collaborations], arXiv:0903.2503; 
T.~Aaltonen {\it et al.}~[CDF Collaboration], 
{\it Phys. Rev. Lett.} {\bf 105}, 252001 (2010).

\bibitem{wmap} E.~Komatsu \etal~[WMAP Collaboration],
\apjs{192}{2011}{18};\\ 
{\tt http://lambda.gsfc.nasa.gov/product/map}.

\bibitem{cdm} M.E.~G\'omez, G.~Lazarides, and C.~Pallis,
\prd{61}{2000}{123512}; \plb{487}{2000}{313}.

\bibitem{nra} T.~Nihei, L.~Roszkowski, and
R.~Ruiz de Austri, \jhep{05}{2001}{063}; \ibid{07}{2002}{024}.

\bibitem{microbsg} G.~B\'elanger, F.~Boudjema, A.~Pukhov, and 
A.~Semenov, {\it Comput. Phys. Commun.} {\bf 174}, 577 (2006).

\bibitem{scn0} G.F.~Giudice, E.W.~Kolb, and A.~Riotto,
\prd{64}{2001}{023508}; N.~Fornengo, A.~Riotto, and S.~Scopel,
\ibid{67}{2003}{023514}; G.B.~Gelmini and P.~Gondolo, 
\ibid{74}{2006}{023510}; G.~Barenboim and J.D.~Lykken, 
\jhep{12}{2006}{005}; A.B.~Lahanas, N.E.~Mavromatos and D.V.~Nanopoulos,
\plb{649}{2007}{83}; M.~Drees, H.~Iminniyaz, and M.~Kakizaki, 
\prd{76}{2007}{103524}.

\bibitem{scn} C.~Pallis, {\it Astropart. Phys.} {\bf 21}, 689 (2004); 
\jcap{10}{2005}{015}; {\it Nucl. Phys.} {\bf B751}, 129 (2006);
hep-ph/0610433.

\bibitem{candidates} K.~Matchev, hep-ph/0402088;  E.A.~Baltz,
astro-ph/0412170; G.~Lazarides, hep-ph/0601016; M.~Taoso, 
G.~Bertone, and A.~Masiero, \jcap{03}{2008}{022}.

\bibitem{gravitino} J.R.~Ellis, K.A.~Olive, Y.~Santoso, and 
V.C.~Spanos, \plb{588}{2004}{7};  
K.~Jedamzik, K.-Y.~Choi, L.~Roszkowski, and R.~Ruiz de Austri, 
\jcap{07}{2006}{007};
J.~Pradler and F.D.~Steffen, \plb{666}{2008}{181}.

\bibitem{axino} L.~Covi, L.~Roszkowski, 
R.~Ruiz de Austri, and M.~Small, \jhep{06}{2004}{003};  
K.-Y.~Choi, L.~Covi, J.E.~Kim, and L.~Roszkowski, 
arXiv:1108.2282.

\bibitem{Baerax} H.~Baer and H.~Summy, 
{\it Phys. Lett. B} {\bf 666}, 5 (2008); 
H.~Baer, M.~Haider, S.~Kraml, S.~Sekmen, and H.~Summy,
\jcap{02}{2009}{002}; 
H.~Baer, A.D.~Box, and H.~Summy, \jhep{10}{2010}{023}.

\bibitem{bsgexp} E.~Barberio \etal~[Heavy Flavor Averaging Group],
arXiv:0808.1297.

\bibitem{bsgSM} F.~Borzumati and C.~Greub, 
{\it Phys. Rev. D} {\bf 58}, 074004 (1998);
M.~Misiak \etal, {\it Phys. Rev. Lett.} 
{\bf 98}, 022002 (2007).


\bibitem{nlobsg} M.~Ciuchini, G.~Degrassi, P.~Gambino, and 
G.F.~Giudice, \npb{527}{1998}{21}; 
G.~Degrassi, P.~Gambino, and G.F.~Giudice,
\jhep{12}{2000}{009}; 
M.E.~G\'omez, T.~Ibrahim, P.~Nath, and S.~Skadhauge, 
\prd{74}{2006}{015015}.

\bibitem{bsmm} P.H.~Chankowski and L.~Slawianowska,
{\it Phys. Rev. D} {\bf 63}, 054012 (2001); 
C.S.~Huang, W.~Liao, Q.S.~Yan, and S.H.~Zhu, 
\ibid{63}{2001}{114021}; {\bf 64}, 059902(E) (2001); 
C.~Bobeth, T.~Ewerth, F.~Kruger, and J.~Urban,
\ibid{64}{2001}{074014}; 
A.~Dedes, H.K.~Dreiner, and U.~Nierste,
{\it Phys. Rev. Lett.} {\bf 87}, 251804 (2001); 
J.R.~Ellis, K.A.~Olive, and V.C.~Spanos, 
{\it Phys. Lett. B} {\bf 624}, 47 (2005).

\bibitem{mahmoudi} F.~Mahmoudi, 
{\it Comput. Phys. Commun.} {\bf 180}, 1579 (2009).

\bibitem{bmmexp} T.~Aaltonen \etal~[CDF Collaboration],
{\it Phys. Rev. Lett.} {\bf 100}, 101802 (2008).

\bibitem{Btn} G.~Isidori and P.~Paradisi, 
{\it Phys. Lett. B} {\bf 639}, 499 (2006); 
G.~Isidori, F.~Mescia, P.~Paradisi, and D.~Temes, 
{\it Phys. Rev. D} {\bf 75}, 115019 (2007).

\bibitem{gmuon} S.P.~Martin and J.D.~Wells,
{\it Phys. Rev. D} {\bf 64}, 035003 (2001).

\bibitem{g2exp} G.W.~Bennett \etal~[Muon $g-2$ Collaboration],
{\it Phys. Rev. D} {\bf 73}, 072003 (2006).

\bibitem{Hagiwara} K.~Hagiwara, R.~Liao, A.D.~Martin, D.~Nomura, 
and T.~Teubner, {\it J. Phys. G} {\bf 38}, 085003  (2011).

\bibitem{lepmh}
The LEP Collaborations ALEPH, DELPHI, L3, OPAL, The LEP Working
Group for Higgs Boson Searches, 
{\it Eur. Phys. J. C} {\bf 47}, 547 (2006).

\bibitem{news1}
O.~Buchmueller \etal, arXiv:1112.3564.

\bibitem{2loops} G.~Degrassi, P.~Slavich, and F.~Zwirner,
{\it Nucl. Phys.} {\bf B611}, 403 (2001);  
A.~Brignole, G.~Degrassi, P.~Slavich, and F.~Zwirner, 
\ibid{B631}{2002}{195}; \ibid{B643}{2002}{79}; 
A.~Dedes, G.~Degrassi, and P.~Slavich, \ibid{B672}{2003}{144}.

\bibitem{comparisons}  B.C.~Allanach, S.~Kraml, and
W.~Porod, \jhep{03}{2003}{045}; 
B.C.~Allanach, A.~Djouadi, J.L.~Kneur, W.~Porod, and 
P.~Slavich, \ibid{09}{2004}{044}.

\bibitem{news2} D.~Albornoz Vasquez, G.~Belanger, R.M.~Godbole, 
and A.~Pukhov, arXiv:1112.2200;  
I.~Gogoladze, Q.~Shafi, and C.S.~Un, arXiv:1112.2206.



\end{thebibliography}
\end{document}